\documentclass[journal]{IEEEtran}
\IEEEoverridecommandlockouts
\usepackage{cite}
\usepackage{amsmath,amssymb,amsfonts}
\usepackage{algorithmic}
\usepackage{graphicx}
\usepackage{textcomp}
\usepackage{float}
\usepackage{graphbox}
\usepackage{mathtools}
\graphicspath{ {IEEEtran/images/} }
\usepackage{booktabs,caption}
\usepackage{wrapfig}
\usepackage{makecell}
\usepackage{dirtytalk}
\usepackage{subcaption}
\usepackage{tabularx,ragged2e,booktabs,caption}
\usepackage[export]{adjustbox}
\usepackage{booktabs,caption}
\usepackage[flushleft]{threeparttable}

\begin{document}
%
\title{Deep-Learning Based Adaptive Ultrasound Imaging from Sub-Nyquist Channel Data}

\title{Deep-Learning Based Adaptive Ultrasound Imaging from Sub-Nyquist Channel Data}
\author{Alon Mamistvalov, Ariel Amar, Naama Kessler and Yonina C. Eldar, \IEEEmembership{Fellow, IEEE}
\thanks{
This work was supported in part by the Igel Manya Center for Biomedical Engineering and Signal Processing, as well as the Benoziyo Endowment Fund for the Advancement of Science, the Estate of Olga Klein -- Astrachan, and the European Union's Horizon 2020 research and innovation program under grant No. 646804-ERC-COG-BNYQ;}

\thanks{A. Mamistvalov, A. Amar, N. Kessler and Y. C. Eldar are with the Faculty of Math and CS, Weizmann Institute of Science, Rehovot, Israel, email:\{alon.mamistvalov, ariel.amar, kessler.naama, yonina.eldar\}@weizmann.ac.il.}

\vspace{-0.5cm}
}

\maketitle

\begin{abstract}
Traditional beamforming of medical ultrasound images relies on sampling rates significantly higher than the actual Nyquist rate of the received signals. This results in large amounts of data to store and process, imposing hardware and software challenges on the development of ultrasound machinery and algorithms, and impacting the resulting performance.
In light of the capabilities demonstrated by deep learning methods over the past years across a variety of fields, including medical imaging, it is natural to consider their ability to recover high-quality ultrasound images from partial data. Here, we propose an approach for deep-learning based reconstruction of B-mode images from temporally and spatially sub-sampled channel data. 
We begin by considering sub-Nyquist sampled data, time-aligned in the frequency domain and transformed back to the time domain. The data is further sampled spatially, so that only a subset of the received signals is acquired. 
The partial data is used to train an encoder-decoder convolutional neural network, using as targets minimum-variance (MV) beamformed signals that were generated from the original, fully-sampled data. 
Our approach yields high-quality B-mode images, with higher resolution than previously proposed reconstruction approaches (NESTA) from compressed data as well as delay-and-sum beamforming (DAS) of the fully-sampled data. In terms of contrast to noise ratio, our results are comparable to MV beamforming of the fully-sampled data, thus enabling better and more efficient imaging than is mostly used in clinical practice today.
\end{abstract}


%
\IEEEpeerreviewmaketitle

\section{Introduction}
\label{sec:introduction}
\IEEEPARstart{O}{ver} the past decades, ultrasound has become a preferred scanning modality in a variety of clinical scenarios due to its non-ionizing and non-invasive nature, high availability and relatively low cost. The scanning process consists of transmitting acoustic pulses to the tissue, receiving their echos and digitally compensating their arrival time due to the geometry, and finally summing over the channel. This process is also referred to as beamforming.


Performing beamforming in the time domain necessitates high sampling rates of the received signals. This requirement originates from the time-alignment step, in which sufficient delay resolution is obtained through over-sampling and interpolation. In practice, signals are sampled at rates 4 to 10 times higher than the transducer central frequency, leading to sampling rates much higher than the Nyquist rate \cite{Steinberg1992, eldar2015sampling, Chernyakova2014}, which is considered as the minimum sampling rate to allow full reconstruction of the continuous-time signal before sampling\cite{Shannon1949, eldar2015sampling}. This leads to vast amounts of samples to transmit and process in otrder to produce the final image.

Data volume is crucial in particular in receive beamforming.
At this stage, averaging the signals across the array is performed using either a pre-defined apodization, as in Delay-and-Sum (DAS) beamforming \cite{Thomenius1996}, or a per-pixel data-adaptive apodization, as in  Adaptive Beamforming using Deep Learning (ABLE) \cite{Luijten2019} or minimum-variance (MV) beamforming \cite{Capon1969}. While the latter allows for better trade-off between the main lobe's width and side-lobes intensity, translating to improved resolution and contrast in the final image, its computational cost is high and increases with data size. Therefore, it is more expensive and harder to implement MV beamforming in real-time clinical applications, making DAS the method of choice, resulting in degraded image quality \cite{Luijten2019}. 


To circumvent the long processing time and the high computational cost, a variety of techniques have emerged which enable reconstruction of the beamformed signal from partial data. In \cite{Chernyakova2014, FourierDomainReconstruction2016}, the authors shift the process of time-alignment to the frequency domain by drawing a connection between the set of Fourier coefficients of the received signals pre-alignment and the set of Fourier coefficients of the beamformed signal, allowing to sample the former at their effective Nyquist rate. A reduction to a sub-Nyquist rate is also considered, by sampling only a subset of the Fourier coefficients of the received signals. Reconstruction is then performed using compressed sensing (CS) methods \cite{eldar2015sampling,eldar2012compressed,CompressedSensing2012}, relying on the Finite-Rate-of-Innovation structure of the beamformed signal \cite{Tur2011, eldar2015sampling, eldar2012compressed}.

In \cite{Cohen2018}, a method is proposed to reduce the data and hardware burden by using sparse arrays, namely, only a subset of the receive elements are activated. Processing is then performed by convolutional beamforming, which is shown to preserve the array beampattern under appropriate conditions on the chosen array \cite{cohen2020sparse}. More specifically, the achieved beampattern is equivalent to that of a virtual array given by the sum co-array of the sparse array. Thus, using a sparse array whose sum co-array includes a full uniform linear array, yields enhanced resolution and contrast from fewer transmitting elements. This method is integrated with Fourier domain beamforming in \cite{Mamistvalov2020,mamistvalov2020compressed}, allowing sub-sampling in both space and time.

The above mentioned works perform recovery from the partial samples by solving a minimization problem in an iterative manner. Since the process is repeated for every acquisition angle in the frame, this results in long processing times. Moreover, fixed weighting is applied on the signals prior to their summation, which results in degraded image resolution in comparison to adaptive beamforming methods \cite{Capon1969,Luijten2019}. To improve on this, one would prefer to replace the fixed weighting with adaptive weights; However, those are calculated per scanned depth and transmission angle, making the process computationally expensive.

Inspired by the notable performance of deep learning over the past years across a wide range of fields and tasks \cite{Lecun2015, Guo2016, Wang2019}, medical imaging included \cite{Greenspan2016, Sahiner2019}, different uses of deep learning in ultrasound reconstruction have been investigated \cite{VanSloun2019}. In \cite{Luijten2019}, a deep-learning based MV beamformer is proposed, implemented with a fully-connected neural network over fully-sampled and spatially sub-sampled channel data. The objective is to ease the heavy computational burden of adaptive beamforming, as well as to improve performance by learning from samples. The works \cite{Khan2019a, Khan2019, Simson2019}  target the same problem using encoder-decoder architectures, while \cite{Zhuang2019} considers the combination of deep neural networks and MV beamforming for contrast enhancement; It suggests an ensemble of networks operating in the frequency domain over frequency sub-bands, either before or after computation of the adaptive weights. 
In \cite{Yoon2017}, the authors expand the problem to sub-sampling of transducer elements and transmission angles. They theoretically justify their approach by drawing a connection between the encoder-decoder architecture and a low-rank Hankel matrix decomposition which models the problem. 
Other works that consider reconstruction from a partial set of plane-wave transmissions use fully convolutional networks \cite{Ghani2019, Gasse2017}, encoder-decoder networks \cite{Simson2018, Senouf2018, Vedula2018a, Perdios2018} and generative adversarial networks (GANs) \cite{Nair2019, Choi2018}.

The aforementioned approaches propose fixed sampling schemes that do not depend on the transmitted pulses or the task at hand. This aspect is addressed in \cite{Huijben2019}, where two concatenated models are proposed. The first model learns to sub-sample the data, while the second, whose architecture depends on the desired task, learns the recovery. Both are jointly trained in an end-to-end fashion.
The method is tested in several different sub-sampling tasks, including temporal sub-sampling of partial Fourier measurements and recovery of the original signal from them. However, this particular task is tested \textit{in-silico} only over simulated random signal vectors; time-alignment of the sub-sampled signals, which is required for ultrasound image recovery, is not addressed, and neither is beamforming. Since traditional time-alignment is not possible over a sub-sampled grid, the application to recovery of temporally sub-sampled ultrasound data remains unexplored. 

To the best of our knowledge, no other deep-learning based method has been proposed for the recovery of temporally partial ultrasound channel data. Here, we address this issue based on the results in \cite{Chernyakova2014} and \cite{Cohen2018}. The input to our model is partial Fourier measurements of the received signals, from either the full array or from a subset of channels, emulating a sparse array. Those can be obtained by sub-Nyquist sampling implemented in hardware, as shown in \cite{Tur2011, Baransky2014}. Using a convolutional neural network (CNN) trained separately on each reduction factor with MV-beamformed targets, our model learns a transformation from the sub-sampled grid to a high quality B-mode image.
Despite the significant reduction in data volume, the CNN outperforms fully-sampled DAS in terms of resolution, and is comparable to fully-sampled MV-beamforming in terms of contrast to noise ratio. As such, it offers an improved image quality compared to the DAS beamformer used in clinical systems today, at a much lower cost in terms of hardware and software

The remainder of this paper is organized as follows: Section \ref{sec:existing} shortly reviews DAS and Fourier domain beamforming. Section \ref{sec:method} introduces our proposed method, which is verified in Sections \ref{sec:expSetup} and \ref{sec:res}. Results are discussed in Section \ref{sec:conc}.

\section{Existing Beamforming Methods}
\label{sec:existing}

\subsection{Delay-And-Sum beamforming}
Consider a phased-array transducer of $M$ elements aligned along the x-axis, where $m_{0}$ denotes the central element, and $\delta_{m}$ denotes the distance to the \textit{m}th element. 

The imaging cycle begins at time $t=0$, when a short pulse is transmitted from the array in direction $\theta$. Denote by $\left(x,z\right)=\left(ct\sin\theta, ct\cos\theta \right)$ the coordinates of the pulse at time \textit{t}, as it propagates through the tissue at speed \textit{c}.
Assume that a point reflector positioned at this location scatters the energy such that an echo is received by all array elements, at a time depending on their location. Beamforming is the operation of averaging the reflections while compensating for these differences in arrival time.

Let $\varphi_{m}\left(t\right)$ be the signal received by the \textit{m}th element, and
\begin{equation}
\begin{aligned}
   \hat{\tau}\left(t;\theta\right) = t + \frac{d_{m}
\left(t;\theta\right)}{c}
\end{aligned}
\end{equation}
be its time of arrival, where
\begin{equation}
\begin{aligned}
   d_{m}\left(t;\theta\right) = \sqrt{\left(ct\cos\theta\right)^2 + \left(\delta_{m} - ct\sin\theta\right)^2}
\end{aligned}
\end{equation}
is the distance traveled by the reflection to the element.
Applying an appropriate delay to the \textit{m}th signal results in its alignment to the origin $m_{0}$:
\begin{equation}
\begin{aligned}
  \hat{\varphi}_{m}\left(t;\theta\right) = \varphi_{m}\left(\tau_{m}\left(t;\theta\right)\right)
\end{aligned}
\end{equation}
where
\begin{equation}
\begin{aligned}
    \tau_{m}\left(t;\theta\right) = \frac{1}{2}\left(t+\sqrt{t^2-4\left(\delta_{m}/c\right)t\sin\theta + 4\left(\delta_{m}/c\right)^2}\right)
\end{aligned}.
\end{equation}
The final beam is derived by averaging the aligned signals received by the entire array:
\begin{equation}
\begin{aligned}
    \Phi\left(t;\theta\right) = \frac{1}{M}\sum_{m=1}^{M}\hat{\varphi}_{m}\left(t;\theta\right).
\end{aligned}
\end{equation}

\subsection{Fourier Domain Beamforming}
\label{subsec:FDBF}

Applying appropriate time delays necessitates high sampling rates, 4-10 times higher than the probe's central frequency. To address this problem, it was shown in \cite{Chernyakova2014} that beamforming can be implemented equivalently in the frequency domain, bypassing the need for oversampling as no shifts are actually performed.

Let $c\left[k\right]$ be the \textit{k}th Fourier series coefficient of the beam $\Phi\left(t;\theta\right)$, which can be expressed as
\begin{equation}
\begin{aligned}
   c\left[k\right] = \frac{1}{M}\sum_{m=1}^{M}\hat{c}_{m}\left[k\right]
\end{aligned}
\end{equation}
where $\hat{c}_{m}\left[k\right]$ is given by
\begin{equation}
\begin{aligned}
   \hat{c}_{m}\left[k\right] = \frac{1}{T}\int_{0}^{T}I_{\left[0,T_{B}\left(\theta\right)\right]}\left(t\right)\hat{\varphi}_{m}\left(t;\theta\right)e^{-i\left(2\pi/T\right)kt}dt.
\end{aligned}
\end{equation}
Here, $I_{\left[0,T_{B}\left(\theta\right)\right]}$ is an indicator function for the beam's support and \textit{T} is defined by the penetration depth of the transmitted pulse.
Following the derivation in \cite{Wagner2012, Chernyakova2014}, $\hat{c}_{m}\left[k\right]$ can be written as
\begin{equation}
\begin{aligned}
   \hat{c}_{m}\left[k\right] = \sum_{n}c_{m}\left[k-n\right]Q_{k,m,\theta}\left[n\right]
\end{aligned}
\end{equation}
where $c_{m}\left[k\right]$ are the Fourier coefficients of the signal received in the $m^{th}$ element with no time-alignment applied to it, and $Q_{k,m,\theta}\left[n\right]$ are the Fourier coefficients of a distortion function $q_{k,m}\left(t;\theta\right)$, that effectively transfers the beamforming delays defined in (4) to the frequency domain. The function $q_{k,m}\left(t;\theta\right)$ depends on the geometry of the array alone, and therefore, its Fourier coefficients can be computed offline and stored in memory.

Since most of the energy of \{$Q_{k,m,\theta}\left[n\right]$\} is concentrated around the DC component, the infinite sum in (8) can be approximated sufficiently with the finite sum
\begin{equation}
\begin{aligned}
   \hat{c}_{m}\left[k\right] \cong \sum_{n\in \nu\left(k\right)}c_{m}\left[k-n\right]Q_{k,m,\theta}\left[n\right]
\end{aligned}
\label{eq:FiniteTimeAlignemnt}
\end{equation}
where $\nu\left(k\right)$ depends on the decay properties of \{$Q_{k,m,\theta}\left[n\right]$\}.
Substituting (9) into (6) yields the beamformed signal in the frequency domain:
\begin{equation}
\begin{aligned}
   c\left[k\right] \cong \frac{1}{M}\sum_{m=1}^{M}\sum_{n\in \nu\left(k\right) }c_{m}\left[k-n\right]Q_{k,m,\theta}\left[n\right].
\end{aligned}
\label{eq:beamformed_FDBF}
\end{equation}
Applying an inverse Fourier transform on \{$c\left[k\right]$\} results in the beamformed signal in time. The relationship \eqref{eq:beamformed_FDBF}, proves that the Fourier coefficients of the beam can be obtained as a linear combination of the Fourier coefficients of the non-delayed received signals. Therefore, it is possible to transfer the process of beamforming to the frequency domain while yielding similar results. 

The required set of Fourier coefficients of the received signals can be obtained in hardware using low-rate sampling, significantly lower than the rate required for time-domain beamforming \cite{Tur2011, Baransky2014}. Further reduction to a sub-Nyquist rate is achieved by obtaining only a subset of the coefficients, resulting in a subset of the beam's coefficients. In this scheme, however, the inverse Fourier transform does not sufficiently recover the beamformed signal in time, and additional methods are required for full recovery. 

As indicated in (\ref{eq:beamformed_FDBF}), Fourier domain beamforming incorporates DAS beamforming of the partial data and thus yields similar resolution; however, how to integrate adaptive beamforming is not clear.
To overcome this challenge, we introduce a deep-learning based approach, substituting adaptive beamforming on the sub-sampled grid by learning a direct transformation to a high-quality B-mode image. 

\begin{figure}
\centering
\includegraphics[width=\columnwidth]{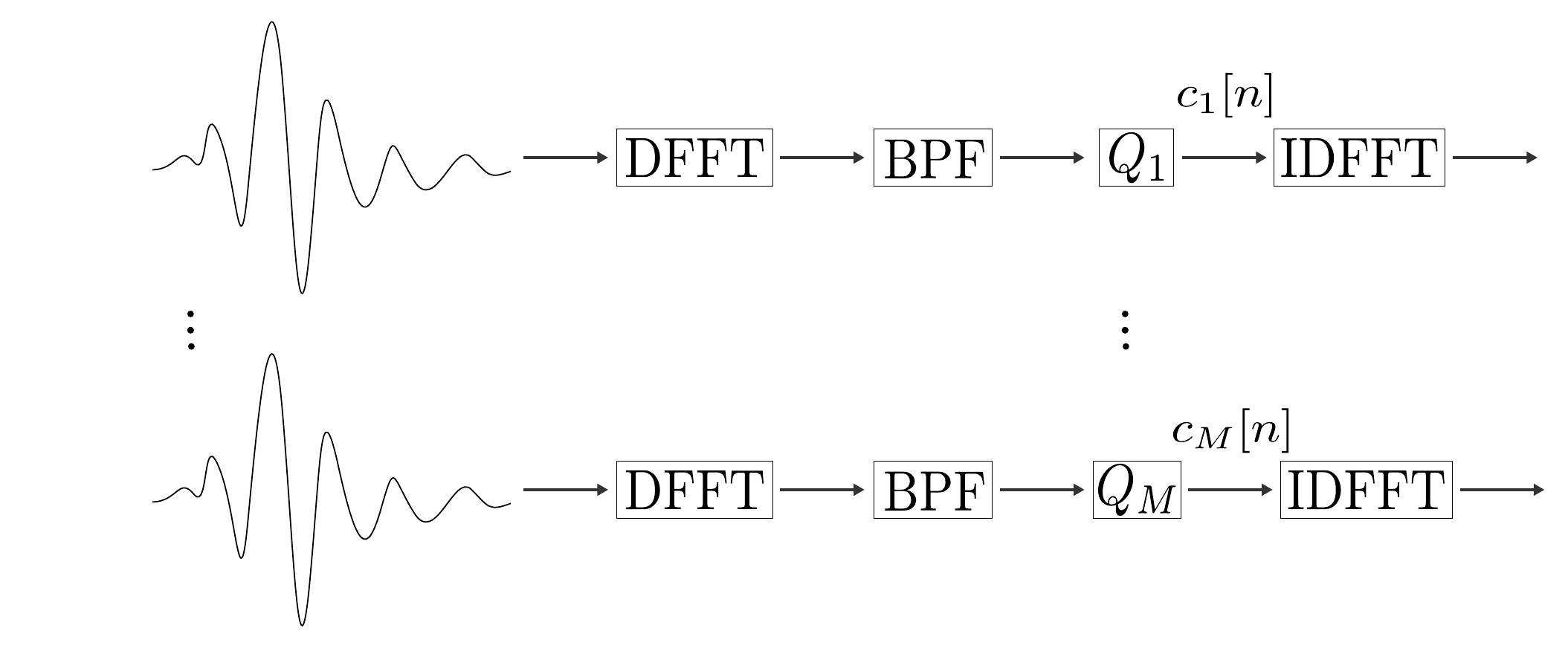}
\caption{Temporal sub-sampling scheme per one transmission angle. Each fully-sampled signal is digitally filtered using a band-pass filter of the desired width around the central transmission frequency. The acquired subset of Fourier coefficients is then multiplied by the corresponding Fourier coefficients \{$Q_{k,m,\theta}\left[n\right]$\} (\ref{eq:FiniteTimeAlignemnt}), effectively applying TOF-correction in the frequency domain \cite{Chernyakova2014}. The resulting signals are zero-padded and transformed back to the time domain.}
\label{fig:temporal_subsampling}
\end{figure}

\begin{figure*}
\centering
\includegraphics[width=\linewidth]{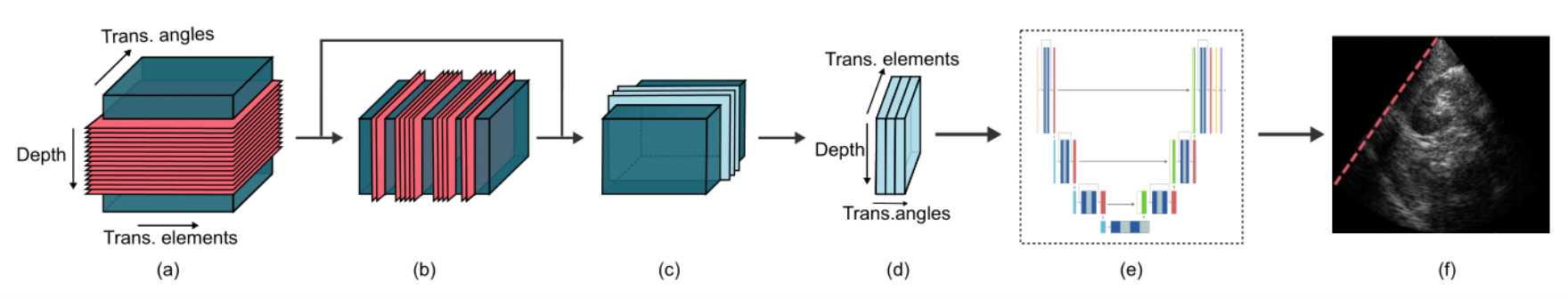}
\caption{Overview of the proposed method's pipeline. (a) Temporal sub-sampling and time-of-flight correction of the fully-sampled data in the frequency domain, as detailed in Fig. \ref{fig:temporal_subsampling}. (b) Spatial sub-sampling, which digitally emulates reception from a sparse array \cite{Cohen2018} (optional). (c) Slicing the sub-sampled data to samples, each consisting of data from three consecutive transmission angles. (d) Permuting the second and third dimensions of each sample, such that the signals received across the array constitute the different channels of input to the network.
(e) Feeding the sample to the neural network. (f) Prediction of the network, which consists of the beamformed radio-frequency signal in one transmission angle.}
\label{fig:data_pipeline}
\end{figure*}

\section{Method}
\label{sec:method}

\subsection{Data sub-sampling and pre-processing}
\label{sub-sampling}

Our pre-processing pipeline consists of digitally emulating temporal sub-sampling with two sampling factors, each constituting a distinct dataset, and applying time-of-flight (TOF) correction in the frequency domain. The two datasets are denoted in the following as $D_{temp_{\times a}}$, $D_{temp_{\times b}}$, where $a,b$ are the sampling factors. Specifically, each dataset is created as follows: The fully-sampled signals are first transformed to the Fourier domain; then, they are filtered using a band-pass filter which corresponds to the desired frequency band around the transmission's central frequency. The resulting coefficients are multiplied with the Fourier coefficients \{$Q_{k,m,\theta}\left[n\right]$\} (\ref{eq:FiniteTimeAlignemnt}), following the method in \cite{Chernyakova2014}. From this point on we assume that we are given only the temporally sub-sampled data. 
Then, the data is transformed back to the time domain by restoring the negative spectrum of each signal (i.e. its transposed-conjugate, since the signals are real-valued), padding with an appropriate-sized vector of zeros to maintain the original resolution in time, and performing an inverse discrete Fourier transform. The process is described in Fig. \ref{fig:temporal_subsampling}.


Since no additional processing steps are performed on the under-sampled data to account for the loss of frequencies outside the selected bandwidth, transforming it back to the time domain introduces aliasing artifacts as well as degraded resolution. However, working in this domain rather than staying in the frequency domain, allows us to model the recovery of the sub-sampled signals and their beamforming as a standard computer-vision problem, somewhat similar to image enhancement or artifact removal. Moreover, it simplifies the task since the difficulties of designing a neural network that operates in the frequency domain and works with complex numbers are avoided.

\begin{figure*}
\centering
\includegraphics[width=\textwidth]{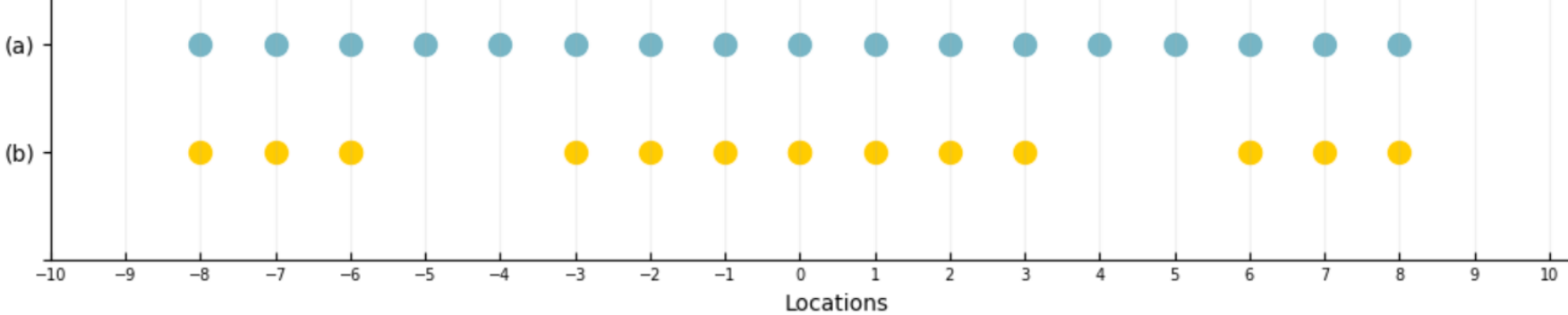}
\caption{Elements position of (a) full array (b) sparse array.
In this example, $M/2 = 9, A = 3$, and $B = 3$.}
\label{fig:spatial_subsampling}
\end{figure*}

To further reduce data volume, a third dataset $D_{spatio-temp}$ is generated by sampling $D_{temp_{\times a}}$ spatially, following the sparse arrays approach presented in \cite{Cohen2018}. The sampling pattern is obtained as follows: 
Let $M$ be the full array's width, and $A, B \in N^{+}$ be a factorization of $M/2$ such that $AB = M/2$. Define the arrays $U_{A}$, $U_{B}$ and $U_{C}$ as
\begin{equation}
\begin{aligned}
    U_{A} =& \{-\left(A-1\right), \dots, 0, \dots, A-1\} \\
    U_{B} =& \{mA:\; m=-\left(B-1\right), \dots, 0, \dots, B-1\} \\
    U_{C} =& \{m:\; |m| = M-A, \dots, M-1\}.
\end{aligned}
\end{equation}
Then the indices of half of the desired array are given by
\begin{equation}
\begin{aligned}
    \hat{U} = U_{A}\cup U_{B} \cup U_{C},
\end{aligned}
\label{eq:SCOBAR_half_array}
\end{equation}
and indices for the entire sparse array are obtained by symmetrically concatenating those around the central element:
\begin{equation}
\begin{aligned}
    U = (M/2 - \hat{U}) \cup (M/2 + \hat{U}).
\end{aligned}
\label{eq:SCOBAR}
\end{equation}
As an example, for $M/2 = 9, A = 3$, and $B = 3$,
the set $\hat{U}$ has only 13 elements out of 17 in the full array, as depicted in Fig. \ref{fig:spatial_subsampling}.

As opposed to \cite{Cohen2018}, no convolution is performed on the filtered data before it is introduced to the network, since such pre-processing resulted empirically in degraded results. This can be explained by the fact that the network does not integrate components inspired by the concept of convolutional beamforming, and moreover, is required to learn different, more complex apodization weights. Since those are computed originally from the signals themselves rather then from the convolved signals, feeding the latter as input to the network might resolve in a harder learning task.



Each sub-sampled dataset is used to train and test the network separately. 
It is introduced to the network in 3D data cubes in order to exploit correlation in all three dimensions, each input sample consisting of all scanned depths per 3 consecutive transmission angles $\theta _{j-1}$,$\theta _{j}$, $\theta _{j+1}$, and all channels of either the original or the sparse array. The network's output is the the beamformed signal at transmission angle $\theta _{j}$  (Fig. \ref{fig:data_pipeline}). Correspondingly, each target consists of the MV-beamformed signal at $\theta _{j}$.

\begin{figure*}
\centering
\includegraphics[width=0.8\linewidth]{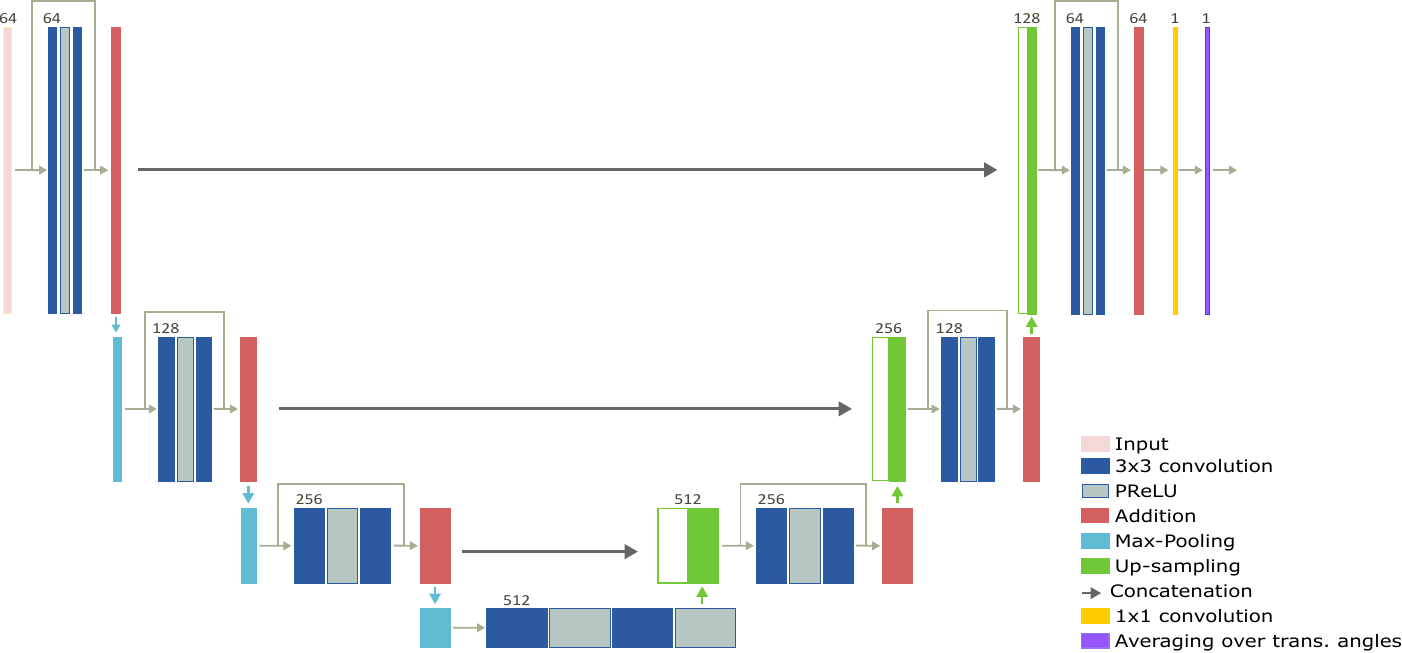}
\caption{Proposed architecture. The number of channels in each block is indicated above the block's first box; the light-gray arrow in each block indicates an additional residual connection, which enables the summation of the block's input with its output. The final averaging of beamformed signals from three transmission angles to one is indicated in the last box.}
\label{fig_net}
\end{figure*}

\subsection{Network architecture}

The desired output of our deep-beamformer is an enhanced B-mode image. Hence, an exact recovery of each sub-sampled signal prior to beamforming is not necessarily required. In other words, we would rather let the network learn a non-exact recovery from the partial samples of each signal to the original signal, if such is able to produce a better beamformed image. To allow that relaxation, we tackle the recovery from partial samples and beamforming jointly in a single model, instead of handling each stage separately.

Our learning model is a variant of UNet, an encoder-decoder CNN which was originally developed for segmentation tasks with limited amount of training data \cite{Ronneberger2015}. 
It consists of three blocks in a contractive path, three blocks in an expansive path and one bottleneck block between them; corresponding blocks are connected with skip connections, i.e. concatenation along the third dimension. Each block of the model consists of two 3x3 convolutions, followed by either a max-pooling layer for a contractive block or an upsampling layer for an expansive block, both operating along the first dimension.
Moreover, Parametric ReLU (PReLU) \cite{Kaiming2018} is used as an activation function rather then ReLU in order to assist the learning process. The model outputs three lines of the beamformed image, which corresponds to the data slice of three consecutive transmission angles we entered as input. We then average over the three angles to get a single line of the beamformed image.

Choosing UNet may seem non-trivial due to the large receptive field created in the deeper layers of the model, towards the bottleneck. It promotes both local and more global features, the later being typically used in tasks which consider large objects, such as segmentation and classification. Nevertheless, UNet's compatibility of to some classes of inverse problems was addressed in \cite{7949028}, and examples of it being applied to different reconstruction problems, ultrasound reconstruction included, were published over the last years \cite{Perdios2020CNNBasedIR, Senouf2018}.
In essence, the class of inverse problems consists of tasks in which a signal is retrieved from a set of measurements, obtained by some forward model. They can be solved by iterative processes which aim to minimize the distance between the actual measurements and the approximated signal after it passes through the forward model, with some regularization derived from prior knowledge about the signal. Such solutions include a multiplication of the operator associated with the forward model with its adjoint and an inversion of the result. Therefore, when the support of the filters associated with these operations is not compact, large receptive field may be a desirable feature in a network that learns to output similar results \cite{7949028}.


\subsection{Loss function}

Ultrasound channel data is characterized by a large dynamic range, and is typically compressed after beamforming to obtain the final B-mode image. Therefore, to promote visual similarity between the output of a learning model and its corresponding beamformed target, we follow \cite{Luijten2019} and apply compression within the loss function used in training, in a variant of mean-squared-error (MSE). The function, named signed-mean-squared-logarithmic-error (SMSLE), is defined as

\begin{equation}
\begin{aligned}
    L_{\textit{SMSLE}} = 0.5 \cdot \| \log_{10}(B_{\mathrm{Pred}}^+) - \log_{10}(B_{\mathrm{MV}}^+) \|_{2}^2  + \\
    0.5 \cdot \| \log_{10}(B_{\mathrm{Pred}}^-) - \log_{10}(B_{\mathrm{MV}}^-) \|_{2}^2
\end{aligned}
\end{equation}
where $B^+$, $B^-$ are the positive and negative parts of the radio frequency beamformed data, $B_{\mathrm{Pred}} $ is the model's prediction and $B_{\mathrm{MV}} $ is the MV-beamformed target \cite{Luijten2019}.


To further promote perception-based similarity between the beamformed signals, a second term, 1D variant of structural similarity index (SSIM) \cite{BREAKEY20133605}, was added to the loss function. This term, which employs similar compression, is given by
\begin{equation}
\begin{aligned}
    L_{\textit{SSIM}} = 0.5 \cdot \left(1-SSIM_{1D}\left(\log_{10}(B_{\mathrm{Pred}}^+), \log_{10}(B_{\mathrm{MV}}^+)\right)\right) + \\
    0.5 \cdot \left(1-SSIM_{1D}\left(\log_{10}(B_{\mathrm{Pred}}^-), \log_{10}(B_{\mathrm{MV}}^-)\right)\right),
\end{aligned}
\end{equation}
where $SSIM_{1D}$ is defined for two signals $x, y$ as
\begin{equation}
\begin{aligned}
    SSIM_{1D}\left( x, y \right)  =  \frac{ \left( 2\mu_{x}\mu_{y} + C_{1} \right) \left( 2\sigma_{xy} + C_{2} \right)}{\left( \mu_{x}^2 + \mu_{y}^2 + C_{1}\right)\left( \sigma_{x}^2 + \sigma_{y}^2 + C_{2} \right)}.
\end{aligned}
\label{eq:SSIM}
\end{equation}
Here, $\mu_{x}, \mu_{y}, \sigma_{x}, \sigma_{y} $ and $\sigma_{xy}$ are the means, standard deviations and cross-correlation of the two images, calculated over an 11-pixel 1D sliding Gaussian window ($\sigma=1.5$), and $C_{1}, C_{2}$ are constants meant to stabilize the division in image regions where the local means or standard deviations are close to zero. We choose the stabilization constants to be $C_{1}=(k_1 \cdot L)^2, C_{2}=(k_2 \cdot L)^2$ where $L$ is the dynamic range of pixel values and $k_{1}=0.01, k_{2}=0.03$ as suggested by Wang \emph{et al.} \cite{Wang2004}. 

Overall, the model was optimized using the following cost function:
\begin{equation}
\begin{aligned}
    L = 0.5\cdot L_{SMSLE} + 0.5\cdot L_{SSIM}.
\end{aligned}
\end{equation}

\begin{figure*}
\centering
\includegraphics[width=0.8\linewidth]{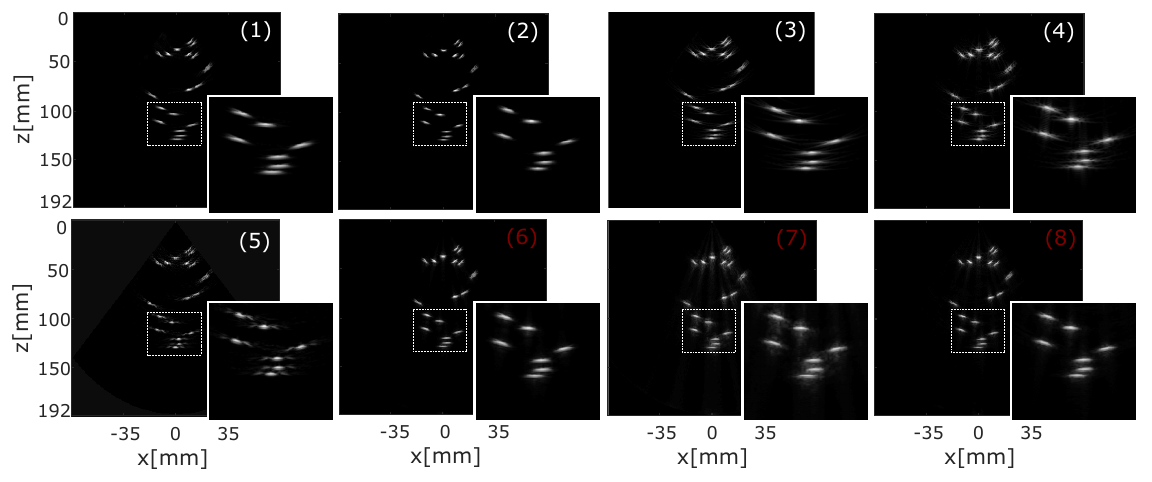}
\caption{Simulated  point scatterers, obtained by (1) DAS (fully-sampled), (2) MV (fully-sampled), (3) NESTA ($D_{temp_{\times5}}$), (4) NESTA ($D_{temp_{\times9}}$), (5) NESTA ($D_{spatio-temp}$), (6) proposed method ($D_{temp_{\times5}}$), (7) proposed method ($D_{temp_{\times9}}$), and (8) proposed method ($D_{spatio-temp}$). In (5), NESTA is applied after convolving the data according to the convolutional beamforming framework. Predictions of the proposed method are averaged over the three models trained per reduction factor. 
While the proposed method demonstrates high axial resolution over the simulated frames, lateral resolution seems slightly degraded, and in frames which were sub-sampled by the x9 sampling factor, axial halos are visually detected.
}
\label{fig:simulated_scatterers}
\end{figure*}

\begin{figure*}
    \centering
      \includegraphics[width=0.8\linewidth]{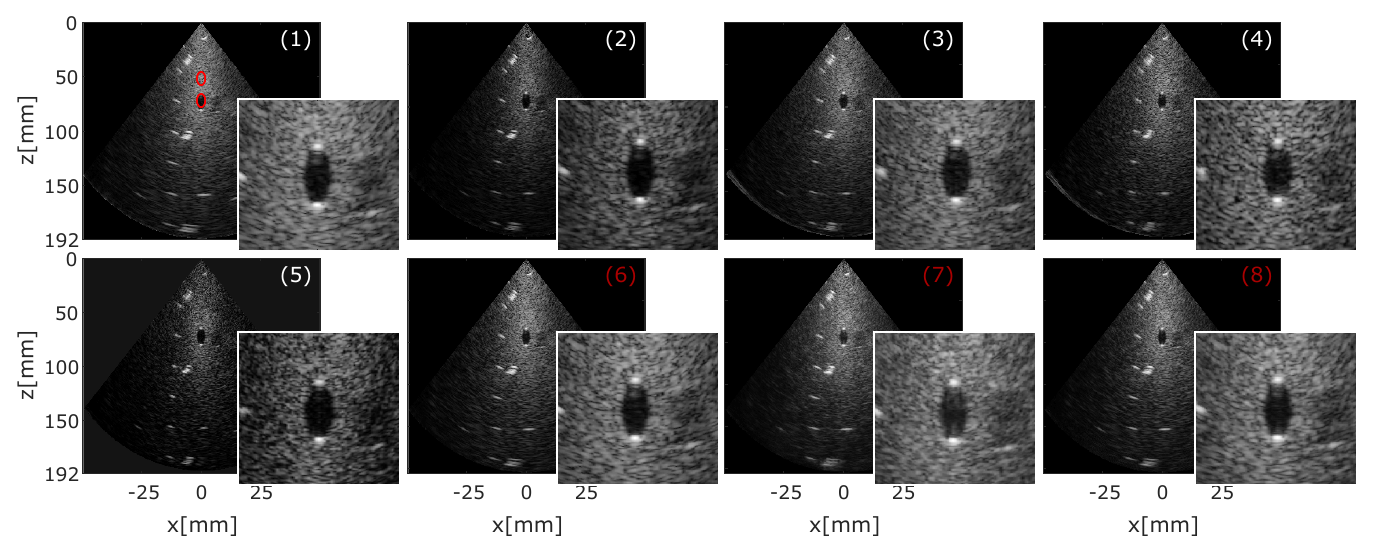}
      \caption{Simulated anechoic cyst, obtained by (1) DAS (fully-sampled), (2) MV (fully-sampled), (3) NESTA ($D_{temp_{\times5}}$), (4) NESTA ($D_{temp_{\times9}}$), (5) NESTA ($D_{spatio-temp}$), (6) proposed method ($D_{temp_{\times5}}$), (7) proposed method ($D_{temp_{\times9}}$) and (8) proposed method ($D_{spatio-temp}$). In (5), NESTA is applied after convolving the data according to the convolutional beamforming framework. Red circles indicate the regions used for computing the CNRs.}
        \label{fig:phantom_contrast}
\end{figure*}
\subsection{Metrics for evaluation}
\label{sec:metrics}

To the best of our knowledge, no other approach was proposed for deep-learning based reconstruction from temporally sub-sampled data. Therefore, we cannot directly compare our model to other deep-leaning based state-of-the-art approaches. Instead, we concentrate on quantitative evaluation to demonstrate the benefits of our model. Contrast-to-Noise-Ratio (CNR) is used to evaluate contrast, and Full-Width-at-Half-Maxima (FWHM), calculated over simulated point scatterers and \textit{in-vivo} data, is used to evaluate axial and lateral resolution. Overall similarity to the target image is evaluated using SSIM.

CNR is evaluated over beamformed phantom scans after envelope detection and logarithmic compression. It is calculated from two regions with different intensities in each image, namely, a simulated cyst and its background, and is given by 
\begin{equation}
  CNR = 20\cdot\log_{10}\left(\frac{\mid \mu_{c}-\mu_{b} \mid}{\sqrt{\sigma_{c}^2 + \sigma_{b}^2}}\right)
\end{equation} 
where $\mu_{c}$, $\mu_{b}$, $\sigma_{c}$ and $\sigma_{b}$ are the means and standard deviations of the cyst and the background, respectively \cite{CNROrigin}.

FWHM is calculated by first computing the score per column and row in each image frame, and then averaging the results over all columns and all rows in the frame, respectively.
SSIM is evaluated per dataset over the test set of each of the three folds, in reference to the fully-sampled MV target images. It is defined for two images $x$, $y$ as in (\ref{eq:SSIM}), with the exception of using a 2D sliding window as in the original definition of the function \cite{Wang2004}.

\section{Experimental Setup}
\label{sec:expSetup}

\textit{In-vivo} data for training and testing was acquired by scanning three healthy volunteers, using a P4-2v Verasonics phased-array transducer with 64 elements. The dataset consists of organs of the abdominal cavity - liver, gallbladder, bladder, kidneys - and the Aorta. The carrier frequency was 2.7 MHz and the sampling rate was 10.9 MHz, which is twice the Nyquist rate, resulting in 1918 samples per image line. Two temporally sub-sampled datasets were generated from it using the described scheme. In the first, $D_{temp_{\times 5}}$, 400 samples per image line were sampled, resulting in sampling rate of 0.42 of the Nyquist rate and 5-fold reduction in comparison to the original data volume. In the second, $D_{temp_{\times 9}}$, 220 samples per image line were sampled, resulting in a sampling rate of 0.23 of the Nyquist rate and 9-fold reduction in comparison to the original data volume. As described in \ref{sub-sampling}, a third dataset, $D_{spatio-temp}$,  was generated by spatial sampling of $D_{temp_{\times 5}}$, omitting 37 out of the 64 transmitting elements. This results in 11-fold reduction in comparison to the original data volume.

Three-fold cross-validation was used in training, resulting in three trained models per dataset. Each model was trained and validated using a different subset of 2 patients in an 80\%-20\% split, setting the third patient aside for testing. 
Each prediction of the model corresponds to the beamformed signal in one transmission angle; as each frame consists from 128 transmissions, this results in $4\cdot 10^{4}$ training samples and $2 \cdot 10^{4}$ testing samples in average per fold.
No organ-wise division was performed during training or testing, meaning that both stages operate on data from multiple organs. Targets for training were generated from MV beamforming of the fully-sampled data, time-aligned in the time domain.

For quantitative evaluation of contrast-to-noise-ratio, tissue mimicking phantoms Gammex 403GSLE and 404GSLE were scanned by the 64-element phased array transducer P4-2v with similar transmission specifications as the \textit{in-vivo} dataset.

The network was implemented with Keras, using Tensorflow backend. It was trained separately on each dataset for up to 100 epochs using Adam optimizer, with an initial learning rate of $3\cdot10^{-5}$ for the temporal sub-sampling and $3\cdot10^{-6}$ for the spatial-temporal sub-sampling. Weights were initialized using He Normal initialization \cite{Kaiming2018}.

\section{Results}
\label{sec:res}

\begin{figure*}
\centering
\begin{subfigure}[b]{0.8\linewidth}
\includegraphics[width=\linewidth]{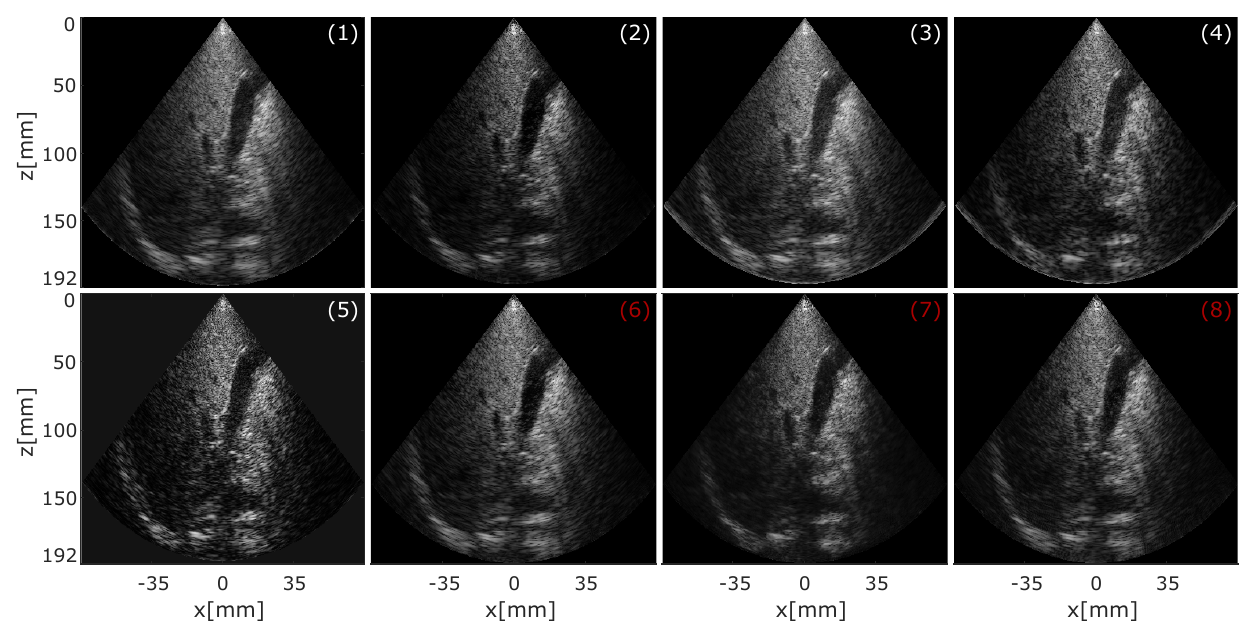}
\end{subfigure}
\begin{subfigure}[b]{0.8\linewidth}
    \includegraphics[width=\linewidth]{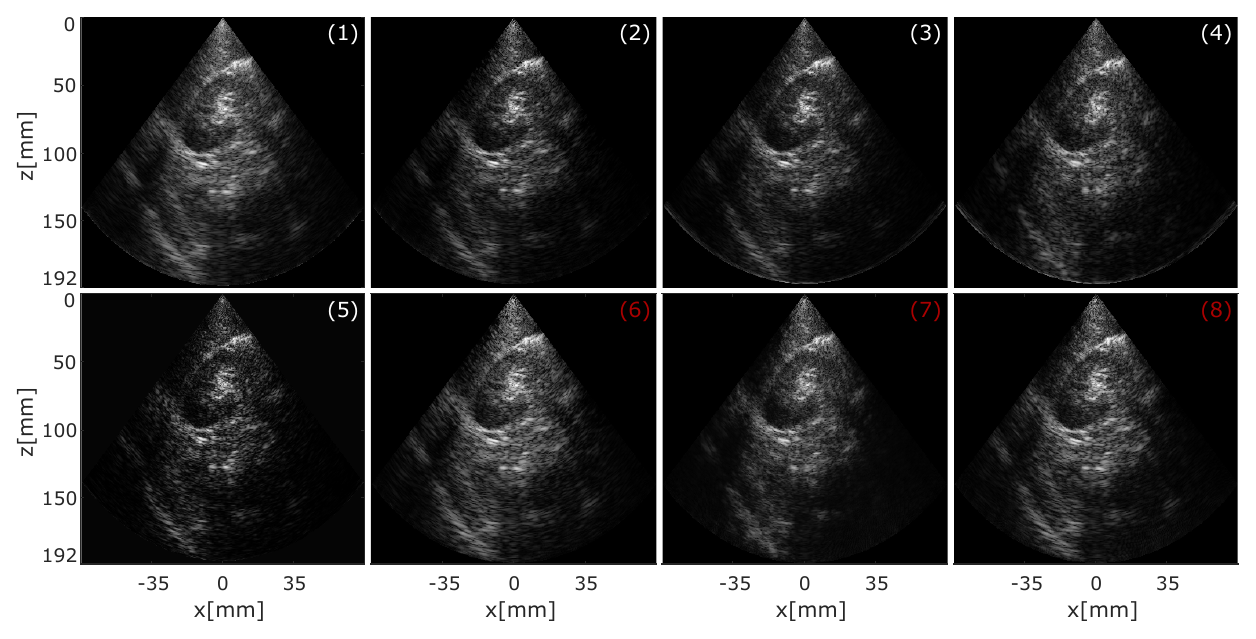}
\end{subfigure}
\caption {Abdominal cavity images from testsets of two of the folds, obtained by (1) DAS (fully-sampled), (2) MV (fully-sampled), (3) NESTA ($D_{temp_{\times5}}$), (4) NESTA ($D_{temp_{\times9}}$), (5) NESTA ($D_{spatio-temp}$), (6) proposed method ($D_{temp_{\times5}}$), (7) proposed method ($D_{temp_{\times9}}$), and (8) proposed method ($D_{spatio-temp}$). In (5), NESTA is applied after convolving the data according to the convolutional beamforming framework.
}
\label{fig:invivo_samples}
\end{figure*}


Table \ref{table:resolution} presents a quantitative evaluation of the lateral and axial resolution of the proposed method, averaged over the three models trained per reduction factor. The results were calculated for simulated point scatterers and \textit{in-vivo} test data, as detailed in Section \ref{sec:metrics}.
As can be seen, our method produces high resolution results on the \textit{in-vivo} data, better than those achieved by the iterative NESTA \cite{Nesterov1983AMF, Becker20111}, used in ~\cite{Chernyakova2014,mamistvalov2020compressed} for reconstruction and in many other CS problems, and better than DAS beamforming of the fully-sampled data. In fact, they are comparable to MV-beamforming of the fully-sampled data. 
High axial resolution is indicated also by the results over the simulated frames. lateral resolution, however, is slightly degraded in comparison to DAS and MV.

An example frame from the simulated set, reconstructed by all compared methods, is presented in Fig. \ref{fig:simulated_scatterers}. There, improved lateral resolution can be seen in the images reconstructed by the suggested method over all reduction factors, in comparison to the NESTA reconstructions. In particular, side beams which are present at the NESTA reconstruction of the x11 reduction factor (Fig \ref{fig:simulated_scatterers}, image 5) are not present at all in the suggested method's reconstruction (Fig. \ref{fig:simulated_scatterers}, image 8). However, in the image which was sub-sampled by the x9 sampling factor (Fig. \ref{fig:simulated_scatterers}, image 7), axial halos are visually detected. We believe that these results stem from the fact that the networks were not trained on synthetic data, and therefore, had a difficulty in compensating for input differences over the temporally-harder dataset.

\begin{table} 
\centering
\caption{Resolution Evaluation with Full Width at Half Maxima (FWHM) Parameter}
\label{table:resolution}
\begin{tabular}{l c c c c c} 
 & \multicolumn{2}{c}{\textbf{Simulated Scatterers}} & \multicolumn{2}{c}{\textbf{\textit{In-Vivo} images}} \\ 
\cmidrule[\heavyrulewidth]{2-5}
 & \textbf{lateral} & \textbf{axial} & \textbf{lateral} & \textbf{axial} \\
\midrule
DAS \textsubscript{fully-sampled} & 0.38 & 0.28 & 0.93 & 0.46 \\ 
\hline
MV \textsubscript{fully-sampled} & 0.3 & 0.28 & 0.68 & 0.32 \\
\hline
NESTA \textsubscript{x5 reduction} & 0.51 & 0.24 & 1.01 & 0.55 \\
\hline
NESTA \textsubscript{x9 reduction} & 0.67 & 0.35 & 1.07 & 0.67 \\
\hline
NESTA \textsubscript{x11 reduction}* & 0.37 & 0.26 & 0.72 & 0.39\\
\hline
\textbf{Proposed \textsubscript{x5 reduction}} & 0.4 & 0.17 & 0.76 & 0.37\\
\hline
\textbf{Proposed \textsubscript{x9 reduction}}  & 0.45 & 0.23 & 0.57 & 0.28 \\   
\hline
\textbf{Proposed \textsubscript{x11 reduction}} & 0.41 & 0.2 & 0.71 & 0.35 \\
\hline
\end{tabular}
\begin{tablenotes}
      \footnotesize
      \item* NESTA\textsubscript{x11 reduction} is applied after convolving the sub-sampled data according to the convolutional beamforming (COBA) framework, as opposed to the proposed method.
      \item** Results are in mm.
    \end{tablenotes}
\end{table}



\begin{table} 
\centering
\caption{Contrast to Noise Ratio (CNR) and Structural Similarity Index (SSIM) Evaluation}
\label{table:CNR_SSIM}
\begin{tabular}{l c c c} 
\textbf{} & \textbf{CNR (dB)} & \textbf{SSIM} \\ 
\cmidrule[\heavyrulewidth]{1-3}
 DAS \textsubscript{fully-sampled}  & 9.99 & 0.75  \\ 
 \hline
 MV \textsubscript{fully-sampled} & 9.31 & 1 \\
 \hline
 NESTA \textsubscript{x5 reduction} & 10.37 & 0.87 \\
 \hline
 NESTA \textsubscript{x9 reduction} & 8.04 & 0.78 \\
 \hline
 NESTA \textsubscript{x11 reduction}* & 7.29 & 0.47\\
\hline
 \textbf{Proposed \textsubscript{x5 reduction}} & 9.82 & 0.75\\
 \hline
 \textbf{Proposed \textsubscript{x9 reduction}} & 8.61 & 0.80 \\   
 \hline
 \textbf{Proposed \textsubscript{x11 reduction}} & 9.47  & 0.77
 \\
 \hline
\end{tabular}
\end{table} 

Table \ref{table:CNR_SSIM} presents a quantitative evaluation of contrast-to-noise-ratio, as well as similarity of the network's outputs to the MV targets, assessed on the phantom scans and \textit{in-vivo} test scans, respectively. Again, results were averaged over the three models trained per reduction factor. As can be seen, our method surpasses DAS's SSIM score at all sub-sampling rates, indicating better resemblance to the fully-sampled MV-beamformed targets. 
However, the performance is slightly degraded in comparison to NESTA. 
In terms of contrast to noise ratio, our method is comparable to DAS and MV over the x5 sub-sampling factor, but is slightly degraded over the x9 sampling factor; in comparison to NESTA, it is again slightly degraded over the x5 sub-sampling factor, but is better over the x9 sampling factor. These last results suggest more robustness to the reduction in sampling rate. 
A comparison view of the beamformed cyst is presented in Fig. \ref{fig:phantom_contrast}, where indeed, an improved contrast in comparison to DAS can be seen in the x5 and x11 reduction factors (images 6, 8), and a degradation is apparent over the x9 reduction factor (image 7).

Fig. \ref{fig:invivo_samples} presents temporally and spatially sub-sampled \textit{In-vivo} data from testsets of the two of the folds, beamformed by the proposed method and compared to the other approaches. Over $D_{temp_{\times5}}$ and $D_{spatio-temp}$, our method yields results that are visually comparable to the target images in terms of both resolution and contrast. Moreover, there is good preservation of speckles and weaker reflections. Over $D_{temp_{\times9}}$, however, slight suppression of speckles and mild pale artifacts are detected, especially in low SNR areas at the margins of the frame.

Finally, we consider two issues regarding the suggested method. First, unlike other adaptive beamforming techniques, such as MV, the proposed method, once learned can be used to beamform different types of scanned objects. On the other hand, a possible drawback of the proposed mechanism can be the fact that it does not produce an actual B-mode US image, but only a single beamformed image line. However, there are well known steps of post-processing to visualize the image which can be added here as well.


\section{Conclusion}
\label{sec:conc}
In this work we presented a deep-learning based method for high quality reconstruction of temporally and spatially sub-sampled channel data, obtained by the schemes presented in \cite{Chernyakova2014} and \cite{Cohen2018}. We have shown that an encoder-decoder CNN, trained on multi-organ scans using a loss function which incorporates domain knowledge, can be used to directly learn the production of high-quality B-mode images from the degraded data.
Our method yields results with high resolution and contrast-to-noise-ratio, which are comparable to model-based beamforming of the fully-sampled data. It performs particularly well on moderate temporal sampling, either with or without spatial sampling; yet, it seems to be more resilient to reduction in temporal sampling rate than the iterative reconstruction method which was previously used for the task in \cite{Chernyakova2014} and \cite{Cohen2018}.
These results indicate that our method can be plugged in to previously proposed schemes of sub-Nyquist and sparse-array ultrasound processing, to improve performance while alleviating the requirements of the data's transmission and processing.

\section*{Acknowledgment}
The authors thank Dr. I. Aharony for performing the volunteer scans which provided the data for training and testing of the proposed method.



%

\newpage
\bibliographystyle{IEEEtran/myIEEEtran}
\bibliography{paper_v2}
\end{document}